\documentclass[prl,twocolumn,showpacs,preprintnumbers]{revtex4}
\usepackage{graphicx,bm}
\begin{document}

\title{Anisotropic Capillary Interactions and Jamming of Colloidal
Particles\\ Trapped at a Liquid-Fluid Interface} 
\author{J.-B. Fournier}
\email{jbf@turner.pct.espci.fr}
\affiliation{Laboratoire de Physico-Chimie Th\'eorique,
E.\,S.\,P.\,C.\,I., 10 rue Vauquelin, F-75231 Paris cedex 05, France}
\author{P. Galatola}
\email{galatola@ccr.jussieu.fr}
\affiliation{LBHP, Universit\'e Paris 7---Denis Diderot, Case 7056, 2
place Jussieu, F-75251 Paris cedex 05, France}
\date{\today}
\begin{abstract}
We determine the capillary attraction and equilibrium configurations of
particles trapped at a liquid-fluid interface due to the pinning of
their contact-line. We calculate analytically the asymptotic interaction
energy between two particles and, numerically, the multi-body energy
landscape for up to four contacting particles. Our results are
consistent with recent experiments.  We show that a system composed of a
large number of such particles behaves as a jammed system.
\end{abstract}
\maketitle

Capillary interactions arise whenever particles that are floating
between a liquid and a fluid distort the interface. These interactions
originate from the dependence of the energy stored in the interface on
the distance between the particles. More generally, interactions
mediated by the elasticity of the medium in which particles are embedded
occur in, e.g., phase-separating binary mixtures~\cite{Beysens85},
critical fluids~\cite{Li92}, bilayer
membranes~\cite{Goulian93,Dommersnes99}, and liquid
crystals~\cite{Ajdari91,Poulin97}. Capillary interactions play
significant roles in several important technologies~\cite{Russel},
including ore flotation~\cite{Wills}, foam stabilization~\cite{Adamson},
and solid surface nanostructuring by deposition of two dimensional
colloidal crystal or
aggregates~\cite{Pieranski80,Fendler96,Burmeister99}.

The interaction between floating objects is well understood only when
the contact-lines can freely adjust their shape at the particles'
surfaces: the weight of the particles induces an interfacial distortion
that relaxes by capillarity, thereby mediating the interaction.  Exact
numerical results are only available, however, for the simplest case of
two parallel cylinders~\cite{Scriven71}. Various approximations, e.g.,
Nicolson's superposition approximation~\cite{Nicolson49}, have been
devised for spheres and
cylinders~\cite{White80,Cloitre88,Kralchevsky93}.

Recently, strong capillary aggregations of floating colloidal spheres
have been reported in a regime where the above gravity-induced
interaction is negligible, due to the smallness of the
particles~\cite{Stamou00}. This effect has been attributed to a
\textit{pinning} of the contact-line at the particles' surface. The
corresponding long-range interfacial distortion has a quadrupolar shape,
which mediates an interaction very strong in units of $k_{\rm B}T$.
Similar behaviors were recently observed in experiments involving
photolithography-fabricated curved discs having an hydrophobic and an
hydrophilic side. It is then the edge of the curved disc that
effectively pins the contact-line~\cite{Brown00}.

In this paper, we model both experiments in an unified way and we
produce exact results for the interaction between such ``trapped''
particles. We obtain the asymptotic interaction \textit{analytically}
without using uncontrolled superposition approximations. We then compute
numerically the exact interaction between two particles, and the
\textit{many-body} interaction between three and four particles, in
order to understand collective behaviors. We finally attempt to connect
our results with the \textit{jammed--fluidized} transition for
attractive particles under external stress~\cite{Liu98,Weitz01}, thereby
proposing these systems as models for micro-rheological studies.

\begin{figure}
\centerline{\includegraphics[width=8cm]{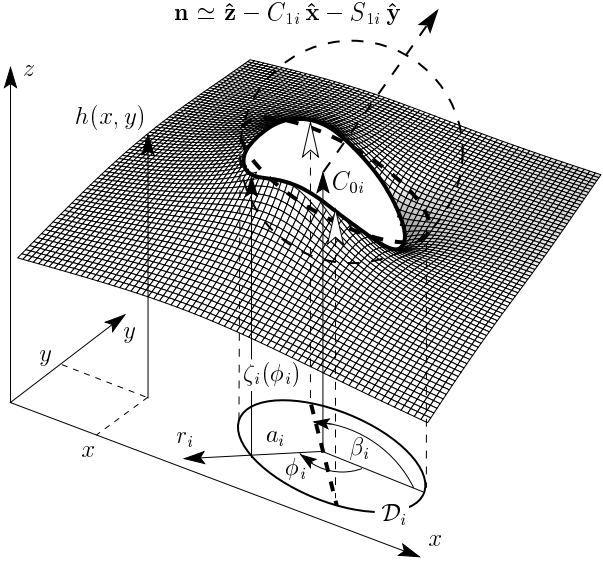}}
\caption{Representation of the contact-line of the $i$-th particle and
of its projection onto the reference plane.}
\label{fig:schema}
\end{figure}

For definiteness, we consider floating particles that posses a fixed,
non-planar line that \textit{pins} the fluid's interface (as in the
above examples~\cite{Stamou00,Brown00}). When such particles interact,
their contact-lines rise, tilt and rotate rigidly with the particles;
however, each line remains fixed in the local frame attached to its
particle. For simplicity, and in agreement with
Ref.~\cite{Brown00}, we assume that the contact-lines depart weakly
from a circle, thus excluding elongated particles. We also assume weak distortions of the
interface height $h(x,y)$ relative to the plane $(x,y)$. This implies
that the projection of each contact-line onto this plane can be
approximated by a circle, even when the particle is tilted. We describe
the $i$-th particle's contact-line by the height function
$\zeta_i(\phi_i)$ (see Fig.\,\ref{fig:schema}), that we expand in
Fourier harmonics: $\zeta_i(\phi_i)=C_{0i}+\sum_n[
C_{ni}\cos(n\phi_i)+S_{ni}\sin(n\phi_i)]$.  Due to the fixed shape of
the contact-line in the particle's frame, some of the above coefficients
are constrained while others are free. $C_{0i}$, which describes how the
particle rises, is free (Fig.\,\ref{fig:schema}). The dipolar
coefficients $C_{1i}$ and $S_{1i}$, which parametrize the tilts of the
contact-line relative to the $z$-axis, are also free. The lowest-order
multipole bearing the information concerning the shape of the
contact-line is therefore the \textit{quadrupole}~\cite{Stamou00}.  We
assume explicitely 
\begin{eqnarray} 
\label{eq:bc} 
\zeta_i(\phi_i)&=&C_{0i}+C_{1i}\,\cos\phi_i +S_{1i}\,\sin\phi_i\nonumber\\
&+&H_{2i}\,\cos\left[2\left(\phi_i-\beta_i\right)\right]
+H_{4i}\,\cos\left[4\left(\phi_i-\beta_i\right)\right].\qquad  
\end{eqnarray}
The quadrupolar coefficient $H_{2i}$ is a \textit{constant}, associated
with the shape of the contact-line, and $\beta_i$ is the free variable
that represents the \textit{rotation} of the particle about its
normal~$\bf n$ (Fig.\,\ref{fig:schema}). For the sake of completeness we
have added the lowest-order multipole that does not break
mirror-symmetry, $H_{4i}\cos[4(\phi_i-\beta_i)]$. The octupolar
coefficient $H_{4i}$ is a \textit{constant} that describes how the
contact-lines departs from a purely quadrupolar shape. The two points
where the height modulation of the contact-line relative to its average
plane is the largest define an \textit{axis}, evidenced by the dashed
line in the figures.

For small deformations, the excess free-energy associated with the
interface's surface tension~$\gamma$ is given by
\begin{equation}\label{eq:F}
F=\frac{\gamma}{2}\int\!d^2r\,
\left(\nabla h\right)^2,
\end{equation}
where ${\bf r}=(x,y)$. For micron-sized particles as we consider here,
gravity is negligible~\cite{Stamou00}. At equilibrium the first
variation of $F$ vanishes, which yields (see Fig.~\ref{fig:schema})
\begin{equation}
\label{var}
\int\!d^2r\left(\nabla^2h\right)\delta h
+\sum_i \oint_{{\cal D}_i}\!ds_i\,\frac{\partial h}{\partial
r_i}\,\delta\zeta_i=0,
\end{equation}
where $ds_i=a_i\,d\phi_i$. Thus $\nabla^2h=0$. For fixed projected
positions and rotations~$\beta_i$ of the particles, $\delta\zeta_i$
reduces to $\delta C_{0i}+\delta C_{1i}\,\cos\phi_i +\delta
S_{1i}\,\sin\phi_i$, since the $\beta_i$'s and the coefficients $H_{2i}$
and $H_{4i}$ in Eq.~(\ref{eq:bc}) are constant. Hence the boundary
equilibrium (freedom of the particles to rise and tilt) is expressed by
the vanishing of the zeroth and first-order Fourier harmonics of
$\partial h/\partial r_i$ on ${\cal D}_i$, which implies that the
expansion of $\partial h/\partial r_i$ starts at $n=2$: 
\begin{equation}
\label{eq:bcd}
\frac{\partial h}{\partial r_i}(r_i\!=\!a_i,\phi_i)=
\!\!\sum_{n=2}^\infty c_{ni}\cos\left(n\phi_i\right)
\!+\!s_{ni}\sin\left(n\phi_i\right).
\end{equation}
The condition $\nabla^2 h=0$ implies the integral equation:
\begin{equation}
\label{eq:int}
h(\mathbf{r}')= \sum_i\oint_{{\cal D}_i} \left[
\frac{\partial h(\mathbf{r})}{\partial r_i} G(\mathbf{r},\mathbf{r}')
-
h(\mathbf{r})\frac{\partial G(\mathbf{r},\mathbf{r}')}{\partial r_i}
\right] ds_i,\!\!
\end{equation}
known as the second Green identity~\cite{Morse},
where $G(\mathbf{r},\mathbf{r}')=\log|{\bf r}-{\bf r}'|/2\pi$ is
$\nabla^2$'s Green function.  Evaluating Eq.~(\ref{eq:int}) for
$\mathbf{r}'$ on the ${\cal D}_i$'s with the Fourier
series~(\ref{eq:bc}) and~(\ref{eq:bcd}), and with the multipolar
expansion:
\begin{equation}
G(\mathbf{r},\mathbf{r}') = \frac{1}{2\pi} \left[
\log r' - \sum_{n=1}^\infty \frac{r^n}{n
r'^n}\cos\left[n(\phi-\phi')\right] \right],
\end{equation}
we obtain a linear set of equations for the unknown variables $C_{0i}$,
$C_{1i}$, $S_{1i}$, and $c_{ni}$, $s_{ni}$ for $n\ge2$.  Given
$\beta_i$, $H_{2i}$ and $H_{4i}$, these variables determine completely
both $h$ and $\partial h/\partial r_i$ on the ${\cal D}_i$'s.  Finally,
Eq.~(\ref{eq:int}) determines $h({\bf r})$ everywhere. The corresponding
free-energy is obtained by integrating Eq.~(\ref{eq:F}) by parts and using
$\nabla^2h=0$: 
\begin{eqnarray} 
F &=& \frac{\gamma}{2}\int\!d^2r\,\nabla\cdot\left(h\,\nabla h\right)=
-\frac{\gamma}{2}\sum_i a_i
\int_0^{2\pi}\!\!d\phi_i\,
\zeta_i\frac{\partial h}{\partial r_i}\nonumber\\
&=&-\frac{\pi\gamma}{2}\!\mathop{\sum_{i;\,k=2,4}} a_i \left[
H_{ki}\left(c_{ki}\cos(k\beta_i)+s_{ki}\sin(k\beta_i)\right)
\right].\qquad
\end{eqnarray}

\begin{figure}
\centerline{\includegraphics[width=7cm]{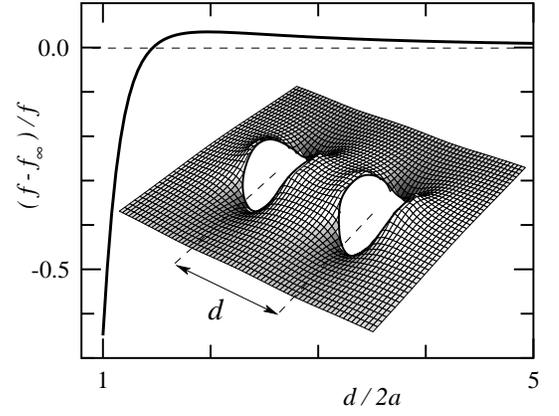}}
\caption{Universal relative error between the exact force~$f$ and the
asymptotic force~$f_\infty$ for two identical, purely quadrupolar
particles ($H_{4i}=0$), of size $a_i=a$, as a function of their reduced
separation.  Inset: calculated interface profile for $\beta_i$'s equal
to zero.}
\label{fig:billes} 
\end{figure}

{\em Two particles.}---Consider two particles $(i=\ell,r)$ separated by
a distance~$d$ (Fig.~\ref{fig:billes}). Solving analytically the linear
system determining the unknown profile coefficients for~$d\to\infty$
yields $F(d)=F_0+F_4(d)+{\cal O}(d^{-6})$, where the leading-order
interaction is
\begin{equation}
\label{eq:F4}
F_4=-12\pi\gamma\frac{{a_\ell}^2{a_r}^2}{d^4}H_{2\ell}H_{2r}
\cos\left[2\left(\beta_\ell+\beta_r\right)\right].
\end{equation}
It coincides with the result found in Ref.~\cite{Stamou00} from a
superposition approximation in the spirit of Nicolson~\cite{Nicolson49}.
Note that it does \textit{not} depend on the $H_{4i}$'s. At fixed $d$,
the particles interaction is minimized when the particles' axes are
symmetrically rotated with respect to the line joining them ($\beta_\ell
= -\beta_r$), \textit{as if the particles were geared}.  Then, the force
acting between them is attractive, asymptotically given by 
\begin{equation}
f_\infty(d)=-48\pi\gamma\frac{{a_\ell}^2{a_r}^2H_{2\ell}H_{2r}}{d^5}.
\end{equation}
The exact interaction energy $\Delta F(d,\beta_\ell,\beta_r)$ can be obtained
for any distance by solving numerically the set of equations for the
profile coefficients after truncating the Fourier expansions at a large
but finite order, and checking for convergence. $f_\infty$ turns to be a
very good approximation for a wide range of separations, as evidenced in
Fig.~\ref{fig:billes}, for two identical particles with $H_{4i}=0$. The
corresponding force at contact is however $\simeq\!65\%$ weaker
than~$f_\infty$. The associated interaction energy is $-2.0\,\gamma
{H_2}^2$, $\simeq\!20\%$ weaker than the approximation given by
$F_4$. For $\gamma\simeq70$\,erg/cm$^2$ (water), $a\simeq1\,\mu$m, and
$H_2/a\simeq0.1$, this gives $\simeq\!-1.4\times10^{-8}$\,erg
($\simeq\!-3\times10^5\,k_{\rm B}T$).  Note that the contact force
actually depends on the details of the higher Fourier components.

\begin{figure}
\centerline{\includegraphics[width=7.5cm]{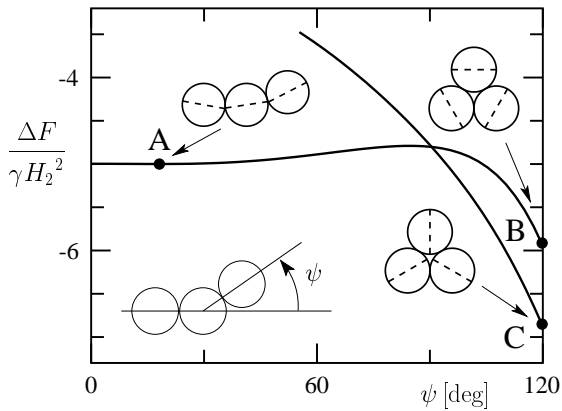}}
\caption{Normalized interaction energy $\Delta F/\gamma {H_2}^2$ for
three identical particles with $H_4=0.25\,H_2$ as a function of the
configuration angle~$\psi$. The two branches come from the minimization
with respect to the angles $\beta_i$. The barrier associated with the
minimum A is $\simeq\!3.6\times10^4\,k_{\rm B}T$ for $H_2=0.1\,\mu$m.}
\label{fig:3}
\end{figure}

When the $H_{4i}$'s are zero (purely quadrupolar particles), the
numerically determined interaction turns out to be proportional to
$\cos\left[2\left(\beta_\ell+\beta_r\right)\right]$ at any
separation: the energy minimum is \textit{degenerate}, corresponding
to $\beta_\ell = -\beta_r$.  Physically, this ``gearing'' condition is
favored because it allows the heights $\zeta_i$ of the particles'
contact-lines to match, thereby preventing a strong local interfacial
distortion~\cite{Brown00}. Our exact calculations further showed that
the presence of nonzero octupolar corrections $H_{4i}$ actually
\textit{lifts} this degeneracy: the minimum energy is attained when
$\beta_\ell=\beta_r=0$ (particles' axes parallel to the separation
vector), as intuited in Ref.~\cite{Brown00}. Indeed, it is best to match
the heights where the contact-lines have a sharper distortion, i.e.,
at the tips of the axes.

{\em Three particles.}---Owing to the strength of the above attraction
($\simeq\!10^5\,k_{\rm B}T$), we study the equilibrium states of three
particles by assuming: i) that each particle contacts at least another
particle, ii) that the particles can rotate around each other, in
agreement with the experiments of Ref.~\cite{Brown00}. We consider
generic particles, i.e., $H_{4i}\ne0$. Without loss of generality, we
fix two particles at contact, and we explore the energy landscape as a
function of the position of the third particle, described by the angle
$\psi$ (Fig.~\ref{fig:3}). For each $\psi$ we minimize the energy with
respect to the rotations $\beta_i$ of the three particles. 

We obtain two distinct branches exhibiting three metastable minima (see
Fig.~\ref{fig:3}).  The latter almost perfectly satisfy the
above-mentioned ``gearing'' rule. The \textit{curved}-line configuration
A suggests a \textit{three-body effect}, since the particle's axes are
not parallel to one-another, contrary to the pairwise case.  Indeed,
repeating the calculation with pairwise interactions instead of the
exact many-body interaction yields a straight line. The most stable,
packed configuration is the triangle C with a radial orientation of the
particles. One might argue that it best relaxes the strong interfacial
distortion located at the facing tips of the particles' axes. The
orthoradial triangle B is also very stable. 

\begin{figure}[t]
\centerline{\includegraphics[width=7.5cm]{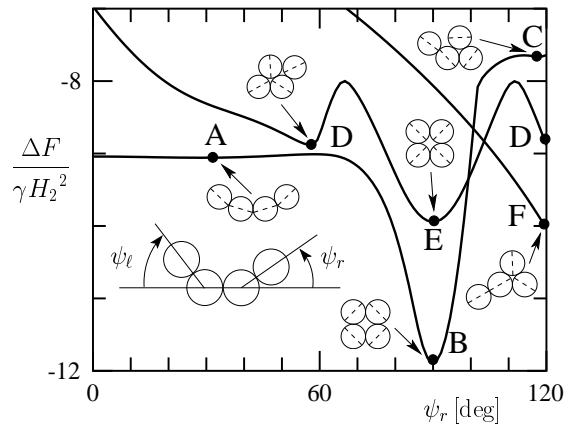}}
\caption{Normalized interaction energy $\Delta F/\gamma {H_2}^2$ for
four identical particles with $H_4=0.4\,H_2$ as a function of the
angle~$\psi_r$ specifying the position of the rightmost particle.}
\label{fig:4b}
\end{figure}

{\em Four particles.}---To determine the equilibrium states of four
contacting particles, we explore the energy landscape by fixing the
positions of three particles through the angle $\psi_r$, and minimizing
with respect to both the position $\psi_\ell$ of the fourth one and all
the rotation angles $\beta_i$ (see Fig.~\ref{fig:4b}).  We obtain three
branches and six equilibrium states. 

As a new indication of a many-body effect, the ``gearing'' rule is
strongly violated in configuration D and slightly violated in C.
Contrary to the three-particle case, the \textit{orthoradial} square~B
is more stable than the radial square~E. One might suspect that since
the gap between the particles is larger than in the triangular case, it
is more important to relax the distortion stored at the axes' tips on
the sides of the square rather than in its center. Note however that
such reasonings become more and more delicate as the number of particles
increases. The fact that B and E are the most stable configurations
suggests, as proposed in Ref.~\cite{Brown00}, that the ground state for
many particles should be a square lattice with~B, or equivalently E, as
unit cell. We find again the curved line A and different triangular
arrangements with a fourth attached particle (C, D, F). Our results do
successfully reproduce the various structures that can be locally seen
in the experimental pictures of Ref.~\cite{Brown00}.

\begin{figure}[t]
\centerline{\includegraphics[width=6.5cm]{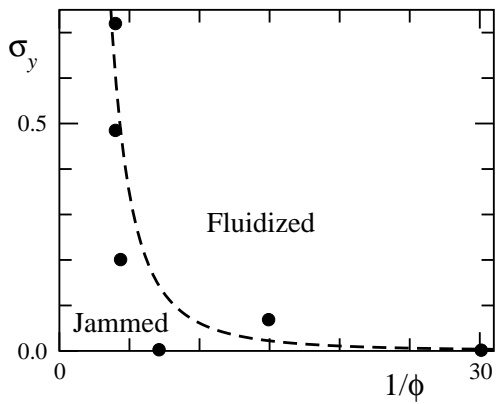}}
\caption{Tentative jamming phase diagram extrapolated from the results
for four particles. The dashed line is the power-law
$\sigma_y\sim\phi^{2.5}$.}
\label{fig:jam}
\end{figure}

Such floating particles can be considered as a model system to study
\textit{jamming} phenomena in complex fluids. Jamming occurs in the
presence of either repulsive or attractive forces~\cite{Weitz01}, when a
disordered system gets trapped in phase space, for kinetic or
energetical reasons, and thus exhibits a yield stress~\cite{Liu98}.
Aggregates of a large number of the ``trapped'' particles discussed in
this paper should form by accretion of individual particles or small
clusters. In the absence of irreversible sticking, at each accretion the
particles will rearrange until they fall in one of the many
metastable configurations. They will thereby form a disordered structure
with a yield stress. Such colloidal aggregates are therefore more akin
to a foam~\cite{Princen83} than to a collection of sticky
spheres~\cite{Doi89}.

To test the relevance of this description in terms of jamming, we
estimate the dependence of the yield stress~$\sigma_y$ on the volume
fraction~$\phi$, from the energetical landscape for four particles
(Fig.~\ref{fig:4b}).  For each metastable state, we set
$\phi\propto1/L^2$, where $L$ is the distance between the farthest
particles, and we estimate $\sigma_y\simeq E_b/aL$, where $E_b$ is the
energy barrier towards the nearest minimum, and $\simeq\!E_b/a$ the
force required to rearrange a given state. This yields the diagram of
Fig.~\ref{fig:jam}, which is quite reminiscent of the phase
diagram for attractive athermal systems proposed in Ref.~\cite{Weitz01},
with a characteristic power-law $\sigma_y\sim\phi^\mu$. 

It should thus be very interesting to experimentally study the
micro-rheological behavior of such particles' aggregates. Indeed, direct
observations would be possible because the system is two-dimensional;
furthermore, as we have shown, this is one of the few systems where
exact many-body calculations are possible.

We thank D. Johannsmann, A. Ajdari, E. Rapha\"el, D. T.  Wu, and P.
Flaud for interesting discussions and comments.

\end{document}